\documentstyle[prl,multicol,aps,epsbox]{revtex}

\begin{document}
\title
{
Physical Origin of the Boson Peak 
Deduced from a Two-Order-Parameter Model of Liquid
}

\author
{ 
Hajime Tanaka}

\address{
Institute of Industrial Science, University of Tokyo, Meguro-ku, 
Tokyo 153-8505
}

\date{Received January 26 2001}
\maketitle
\begin{abstract}
We propose that the boson peak originates from the (quasi-) localized 
vibrational modes associated with long-lived locally favored structures, 
which are intrinsic to a liquid state and are 
randomly distributed in a sea of normal-liquid structures. 
This tells us that the number density of locally favored structures 
is an important physical factor determining 
the intensity of the boson peak. 
In our two-order-parameter model of the liquid-glass transition, 
the locally favored structures act as impurities disturbing crystallization 
and thus lead to vitrification. 
This naturally explains the dependence of the intensity 
of the boson peak on temperature, pressure, and fragility, and also 
the close correlation between the boson peak and 
the first sharp diffraction peak (or prepeak). 

\end{abstract}

%\kword{boson peak, medium-range order, glass transition}

%\sloppy

%\newpage

\begin{multicols}{2}

The dynamics of glass-forming liquids has been extensively studied 
in order to reveal the physical mechanism of the liquid-glass transition. 
One of the most mysterious dynamic modes, which universally 
exists in many different types of glasses, is the low-frequency 
vibrational motion giving rise to the so-called ``boson peak'' 
in the inelastic scattering of light or neutrons \cite{Jackle}. 
The boson peak originates from low-energy excitations 
having the energy of approximately 0.2-2 THz $\cong$ 10-100 K and reflects 
the excess density of vibrational states (DOS) in comparison 
with the Debye value. This excess DOS is, thus, also the origin of 
the low-temperature anomaly of specific heat $C_P$, 
namely, a hump in $C_P/T^3$, which is again universally 
observed for different kinds of glasses. 
The physical origin of the low-energy excitations, or the 
low-temperature anomalies, remains to be clarified in the physics 
of condensed matter. 

The anharmonic contribution below 2 K was, on the other hand, 
successfully explained by two-level-systems models at least 
on a phenomenological level \cite{Anderson,Phillips}, 
although the microscopic origin of 
a two-level-systems feature and its universality are still 
not clear. The connection of this behavior to 
the excess vibrational states is suggested by the soft potential 
model, based on the concept that both originate from 
the soft anharmonic nature of the potentials \cite{Soft1,Soft2,Soft3}. 
Accordingly, the model predicts a close relationship between them. 
However, recent studies by Sokolov et al. \cite{SokolovCp} 
indicate the absence of such a correlation. 
A number of other theoretical models 
have also been proposed to explain the anomalous low-energy excitations 
commonly observed in glasses; e.g., the localization of vibrations due to 
homogeneous disorder whose spatial correlation extends to a medium range 
\cite{Elliott0,Wagner} 
and the frequency resonance or localization of phonons due to 
nano-inhomogeneities (clusters or blobs) 
\cite{Malinovsky,Duval,Schober,Pang,Uchino1,Uchino2}. However, there has so 
far been no consensus on the physical origin of the boson peak. 

The fundamental questions associated with the boson peak, 
or the low-temperature anomalies, are as follows: 
(i) Why is the boson peak frequency $\omega_{\rm bp}$ 
several times smaller than the Debye frequency $\omega_{\rm D}$ 
of the system? What is the origin of such soft excitations? 
(ii) Is the anomaly produced by disorder effects? (iii) Is it 
related to the glass transition itself? 
(iv) How universal is it to glassy disorder systems? 
(v) Is it specific only to a supercooled and glassy state or 
common to a liquid state? 
To answer all these questions, 
we need to clarify the true physical origin of a boson peak.  
One of the difficulties in specifying the mechanism of 
a boson peak is that the relationship between a boson peak and a 
liquid-glass transition itself is not clear. 

In this Letter, we present a natural and coherent physical 
explanation for the boson peak and its connection to 
glass-transition phenomena in the light of our two-order-parameter 
model of a liquid 
\cite{HTglass1,HTglass2,HTglass3,HTglass4,HTglass5,HTW1,HTW2,HTW3}. 
We place a special focus on the experimental finding 
that {\it a boson peak can exist even above the melting point 
$T_m$} for many strong glass formers 
such as SiO$_2$, B$_2$O$_3$, ZnCl$_2$, and glycerol 
\cite{Ruffle,Fytas1,Fytas2,Quitmann}. 
This means that {\it a boson peak can exist even in an equilibrium 
liquid state for some strong liquids}.  

First we briefly summarize our two-order-parameter 
model of a liquid 
\cite{HTglass1,HTglass2,HTglass3,HTglass4,HTglass5,HTW1,HTW2,HTW3}. 
Our model is based on the concept that there are generally 
two types of local structures in a liquid, normal-liquid structures 
and locally favored structures, reflecting complex 
many-body interactions. 
The former is favored  by density order parameter $\rho$, 
which attempts to maximize the local density and leads 
to crystallization, while the latter is favored by 
bond order parameter $S$, which originates from 
the symmetry-selective parts of the interactions and attempts to maximize 
the quality of bonds. 
Locally favored structures are icosahedra for metallic glass formers, 
and tetrahedra and their ring organization 
for covalent-bonding glass formers. 
The locally favoured structure may have the local symmetry of 
a crystal, which is different from that 
of the equilibrium crystal, for a system with polymorphism. 
Thus, we view a liquid as follows: locally favored structures 
with medium-range order are randomly distributed in a sea of 
normal-liquid structures [see Fig. 1 (a)]. 
It should be noted that a locally favored structure has its own specific 
packing symmetry. 
A simple thermodynamic model of the two states with different degeneracies 
\cite{HTglass1,HTglass2,HTglass3,HTglass4,HTglass5,HTW1,HTW2,HTW3} 
tells us that the average fraction of locally 
favored structures, $\bar{S}$, is given by 
\begin{equation}
\bar{S} \sim \frac{g_S}{g_\rho} \exp[\beta(\Delta E-P\Delta v)], 
\label{eq:Sdef}
\end{equation}
where $\beta=1/k_{\rm B}T$ ($k_{\rm B}$: Boltzmann's constant), 
$\Delta E$ and $\Delta v$ are the energy gain 
and the volume increase upon the formation 
of a locally favored structure, respectively, 
and $P$ is the pressure. 
%Note that $\Delta E>0$ and $\Delta v>0$. 
Here $g_S$ and $g_\rho$ are the degrees of degeneracy of the states of 
locally favored structures and normal-liquid structures, respectively. 
We assume that $g_S \ll g_\rho$. This is the direct consequence of 
the uniqueness of locally favored structures and the existence of 
many possible configurations of normal-liquid structures. Thus, there is 
a large loss of entropy upon the formation of a locally favored structure, 
which is given by $\Delta s=k_{\rm B} \ln (g_S/g_\rho)$. 
The validity of eq. (\ref{eq:Sdef}) was confirmed by the successful 
physical description of water's anomalies in terms of our model 
\cite{HTW1,HTW2,HTW3}. 
The lifetime of a locally favored structure can be estimated as 
$\tau_{\rm LFS}=\tau_0 \exp(\beta \Delta G)$, 
where $\tau_0$ is the inverse of the attempt angular 
frequency and $\Delta G$ 
is the energy barrier to be overcome upon the transformation from a 
locally favored structure to a normal-liquid structure.

Since the symmetry of locally favored structures is usually not consistent 
with that of the equilibrium crystal, 
locally favored structures 
act as impurities for density ordering, namely, crystallization. 
Thus, we can apply the knowledge of random-spin systems to this problem. 
Due to the random disorder effects of locally favored structures, 
a liquid enters into the Griffiths-phase-like frustrated state below 
$T_m^\ast$, which is the melting point of the corresponding 
pure system and should be located near the real melting point $T_m$ 
of the material (note that $T_m$ is usually located 
near the so-called mode-coupling $T_c$ 
\cite{HTglass1,HTglass2,HTglass3,HTglass4}). 
Namely, below $T_m^\ast$ the free energy of a system 
starts to exhibit a complex multi-valley structure. 
This leads to the non-Arrhenius behavior of the structural 
relaxation. The Vogel-Fulcher temperature $T_0$ is then 
defined as a transition temperature from 
the Griffiths-phase-like state to the spin-glass-like nonergodic 
state. 
%We stress that this prediction is specific in the sense that $T_m^\ast$ 
%is directly related to the real melting point $T_m$, 
%which is an intrinsic physical property of the material. 

This physical explanation provides a satisfactory answer to the question 
of which physical factor governs the fragility of liquid 
\cite{HTglass1,HTglass2,HTglass3,HTglass4}: 
In our model, fragility is determined by the strength of random 
disorder effects, namely, $\bar{S}$. Thus, a liquid with larger $\bar{S}$, 
which suffers from stronger disorder effects, should be ``stronger'', or 
``less fragile''. This leads to a larger distance between $T_m^\ast$ and 
the Vogel-Fulcher temperature $T_0$ for a stronger liquid, 
which is consistent with experimental results 
\cite{HTglass1,HTglass2,HTglass3,HTglass4}. 
It is worth noting that the only way for a liquid to eliminate these 
random disorder effects, 
is ``global crystallization''. While a system is in a liquid state, it 
is inevitably under the influence of frustration effects. 

\begin{minipage}{8.5cm}
\begin{figure}
\begin{center}
\psbox[width=8.5cm]{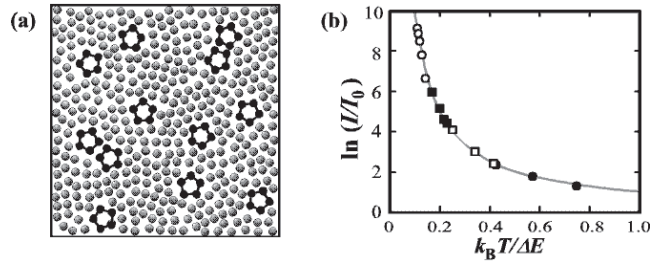}
\end{center}
%\vspace{-0.2cm}
\caption{(a) Schematic snapshot image of a liquid. 
The black particles connected by bonds are participating locally 
favored structures, 
while all the other particles belong to normal-liquid 
structures. Note that the structure is rapidly fluctuating with time. 
%The number density of locally favored structures 
%increases with decreasing $T$ in proportion to $\bar{S}$. 
(b) $T$-dependence of the boson peak intensity above $T_g$, which 
is obtained by subtracting the quasi-elastic component from the 
total scattering. 
The raw boson peak intensity should be proportional to 
$\omega[n(\omega)+1]\bar{S}$. 
Thus, the function $I=I_0 \exp(\Delta E/k_{\rm B}T)$ 
was fitted to the boson peak intensity scaled by $\omega[n(\omega)+1]$ 
by using $I_0$ and $\Delta E/k_{\rm B}$ as adjustable parameters. 
Then $\ln (I/I_0)$ is plotted against 
$\Delta E/k_{\rm B}T$. $\Delta E/k_{\rm B}$=3981 K, 3405 K, 1238 K, and 774 K, 
for As$_2$O$_3$ (open circles), Na$_{0.5}$Li$_{0.5}$PO$_3$ 
(filled squares), B$_2$O$_3$ (filled circles), and glycerol (open squares), 
respectively. 
The data are taken from Refs. 24-27. 
The solid line is the theoretical prediction [see eq. (1)]. 
Note that below $T_g$, $\bar{S}$ should become constant. Thus, the boson peak 
intensity scaled by the temperature bose factor becomes constant 
below $T_g$ in our model, in agreement with the well-known experimental 
finding. 
}
\label{fig:fig1}
\end{figure}
\end{minipage}

\bigskip

On the basis of this model, we seek the physical origin of 
the boson peak. We assign the boson peak to the (quasi-) localized 
cooperative vibrational 
modes associated with a locally favored structure and/or a cluster 
made of a few locally favored structures. Locally favored 
structures are characterized by a particular 
spatial symmetry which is not consistent with the symmetry 
of the ``equilibrium'' crystal, and are distributed randomly in space. 
Their lifetime $\tau_{\rm LFS}$ is much longer than that of 
normal-liquid structures, which are in a disordered state. 
Thus, a locally favored structure can have the cooperative vibrational 
eigen mode with an angular frequency of $\omega_{\rm bp}$ 
unique to that structural unit if $\tau_{LFS}$ is longer 
than $1/\omega_{\rm bp}$. 
This condition determines the high-temperature stability limit 
of the boson peak, $T_{\rm bp}$. 
$\omega_{\rm bp}$ can roughly be estimated as $\omega_{\rm bp} \sim 
2 \pi v_t/R$, where $v_t$ is the velocity of the transverse 
sound and $R$ is the average characteristic length of 
a locally favored structure and a cluster made of a few locally 
favored structures. 
It is worth mentioning here that Uchino and Yoko \cite{Uchino1,Uchino2} 
specified the modes characteristic of particular types of 
medium-range ordering regions, which should correspond to our 
locally favored structures, for B$_2$O$_3$ and SiO$_2$ 
by {\it ab initio} molecular orbital calculations. 

In the light of this explanation, we now consider 
the questions (i)-(v) (see the introduction). The answers are as follows: 
(i) A locally favored structure with medium-range order and its 
cluster are responsible 
for the boson peak and the low-temperature anomaly. 
Their characteristic size, $R$, is 
larger than that of the intermolecular bond, $a$. 
Thus $\omega_{\rm bp}$ is several times smaller than $\omega_{\rm D}$. 
(ii) Locally favored structures and clusters are isolated locally 
ordered structures [see Fig. 1 (a)] 
and are not an average structure of supercooled liquid or glass, which 
is usually described by the term ``medium-range order''. 
Thus, it may not be appropriate to simply say that the boson peak is 
due to disorder effects. (iii) It is a locally favored 
structure that causes frustration disturbing crystallization and 
is thus responsible for vitrification. In our model, thus, the boson peak 
is crucially related to the glass-transition phenomena. 
(iv) The boson peak should universally exist in any glass-forming liquid 
in principle since in our two-order-parameter model of the liquid-glass 
transition \cite{HTglass1,HTglass2,HTglass3,HTglass4} 
it is the existence of locally favored 
structures that is the primary cause of vitrification. 
(v) Locally favored structures, which are responsible for the boson peak, 
exist not only in a supercooled or glassy state, but 
also in a liquid state if $T_{\rm bp}>T_m$. 

Here we summarize important experimental findings, which any physical 
model of the boson peak has to explain and thus can be used as 
criteria for determining the validity of a model. 
(a) The boson peak and the low-temperature anomaly of $C_P$ are 
more pronounced in stronger glass formers 
\cite{SokolovCp,SokolovBP1,SokolovBP2,Ngai}. 
(b) The boson peak intensity \cite{note} 
becomes weaker with increasing either temperature 
or pressure \cite{Inamura,Yagi}. 
(c) A boson peak has its own unique instability point, $T_{\rm bp}$, 
which accompanies the appearance of quasi-elastic contributions. 
It is often located above the melting point $T_m$ for strong glass formers 
\cite{Ruffle,Fytas1,Fytas2,Quitmann}, 
as confirmed in SiO$_2$, B$_2$O$_3$, ZnCl$_2$, and glycerol. 
These results seem to be contrary to 
the conjecture that $T_{\rm bp} \sim T_c$ 
\cite{SokolovTc,Gotze}, although we cannot completely deny the possibility 
that they are exceptional cases. 
More importantly, they indicate that the boson peak can exist 
{\it even in an equilibrium liquid state} above 
$T_m$ (i.e., $T_{\rm bp}>T_m$). 
This cannot be explained by conventional models since 
they assume that an equilibrium liquid is completely 
homogeneous and has no specific structures, while a supercooled 
liquid has some medium-range order or specific disorder 
which is considered to be responsible for the boson peak. 
(d) There is much evidence \cite{Novikov,SokolovFSDP,Sugai,Hemley} 
which suggests the correlation between the boson peak and the first sharp 
diffraction peak (FSDP) (or prepeak) \cite{Elliott}. 
Although there are experimental results \cite{Borjesson} contrary to 
this prediction about the correlation \cite{Novikov}, it may be true 
that the boson peak and the FSDP share some common origin in many systems. 
This is also supported by the fact that the FSDP 
has many features 
similar to the boson peak; for example, it often survives above $T_m$, 
and its intensity decreases with increasing either temperature or pressure 
\cite{Alba}. 

An explanation of all the above-described findings (a)-(e) 
in a coherent manner at least on a qualitative level 
should be the first step toward an understanding of  
the physical mechanism of the boson peak. 
Thus, we attempt to define a coherent physical scenario rather than to 
quantitatively explain a particular experimental finding. 
Hereafter, we evaluate the validity of our explanation in terms of 
the above criteria (a)-(e). 
(a) According to our model, the magnitude of excess DOS, 
which is correlated with the intensity of a boson peak 
and the low-temperature anomaly of $C_P$, should be proportional 
to the number density of the locally favored structure, $\bar{S}$. 
The two-order-parameter model of liquid 
\cite{HTglass1,HTglass2,HTglass3,HTglass4} 
indicates that stronger liquids suffer from stronger disorder effects, which  
are characterized by larger $\bar{S}$. 
Thus, the criterion (a) is naturally satisfied in our model. 
In our model \cite{HTglass1,HTglass2,HTglass3,HTglass4,HTglass5}, 
locally favored structures can be regarded as impurities disturbing 
crystallization 
and this view is consistent with a recent conjecture by Quitmann and 
Soltwisch 
\cite{Quitmann} that the medium-range order responsible for a boson peak 
disturbs crystallization. 
(b) Our model indicates that the intensity of a boson peak \cite{note} 
may be constant 
below $T_g$, but should decrease in proportion to the fraction of locally 
favored structures, $\bar{S}$, in a liquid state above $T_g$, 
with increasing temperature. 
Figure \ref{fig:fig1}(b) shows the $T$-dependence of the boson peak 
intensity for four strong glass formers, which is reasonably 
explained by our prediction. 
The boson peak intensity should also decrease with increasing pressure 
in proportion to $\bar{S}$ above $T_g$. However, there have 
been few experiments on the pressure effects in a liquid state. 
On the other hand, there are many experiments in a glassy state 
below $T_g$, which 
clearly demonstrate the decrease of the boson peak intensity upon 
densification (see, e.g., Refs. 33 and 34). 
Such behavior is consistent with our model, but 
it may not be described by the thermodynamic relation, eq. (\ref{eq:Sdef}), 
in an exact sense because of the intrinsically nonequilibrium nature of 
the densification process. 
Detailed studies on the temperature and pressure dependencies 
of the boson peak intensity ``above $T_g$'' 
are highly desirable to more unambiguously confirm our prediction that 
the boson peak intensity is proportional to $\bar{S}$ between 
$T_{\rm bp}$ and $T_g$. 
We could determine $\Delta E$, $\Delta s$, and $\Delta v$ from such analyses. 
(c) In our model, a boson peak becomes unstable for 
$\omega_{\rm bp}\tau_{\rm LFS} \leq 1$. 
The temperature of the instability of a boson peak, $T_{\rm bp}$, 
is determined as 
\begin{eqnarray}
T_{\rm bp} \sim -\Delta G/[k_{\rm B} \ln(\omega_{\rm bp} \tau_0)]. 
\end{eqnarray}
Above $T_{\rm bp}$, where $\tau_{\rm LFS} <1/\omega_{\rm bp}$, 
the vibrational modes characteristic 
of locally favored structures should be overdamped. 
It is natural to expect a correlation between $\Delta E$ (the strong nature 
of a liquid) and $\Delta G$. 
Thus, a strong glass former with large $\Delta G$ should have 
a high $T_{\rm bp}$ (sometimes above $T_m$). 
This is consistent with the observed behaviors 
\cite{Ruffle,Fytas1,Fytas2,Quitmann}. 
Our model can also naturally explain the existence of the boson peak 
in a liquid state above $T_m$. 
A fragile glass former with small $\Delta G$, on the other hand, 
should have a low $T_{\rm bp}$. For a very fragile liquid, thus, the boson 
peak may not be observed in the liquid state. 
The fact that a boson peak survives above $T_m$ rules out 
the possibility that it is due to small clusters of the equilibrium 
crystal. 
(d) We propose that the FSDP or prepeak observed 
in the structure factor $S(q)$ should reflect locally favored structures 
and their aggregates or clusters, in many cases. 
Recently, Massobrio et al. \cite{Massobrio} studied FSDP in GeSe$_2$ 
by comparing their first-principles molecular dynamics simulations with 
neutron structure factors and found that the FSDP originates from 
the medium-range order of tetrahedral symmetry, GeSe$_4$, and 
its intensity decreases with increasing temperature. 
This is consistent with our explanation. 
Furthermore, Dzugutov et al. \cite{Dzugutov} demonstrated 
that the FSDP is due to well-defined atomic-scale voids, which is 
associated with icosahedral local order for a simple monoatomic liquid. 
On noting that the locally favored structures for such a liquid 
are icosahedral structures, their finding is consistent with 
our explanation that the FSDP reflects the pseudo-periodic atomic density 
fluctuations associated with locally favored structures and their 
clusters. 
This explanation naturally leads to the conclusion that there 
should be a strong correlation between 
the boson peak intensity and the intensity of the FSDP (or prepeak). 
Indeed, this is consistent with the observations 
\cite{Novikov,SokolovFSDP,Sugai,Hemley}. 
Furthermore, this is consistent with the finding that in many cases 
the boson peak intensity and the intensity of the FSDP (or prepeak) 
\cite{Alba} 
both decrease with increase of either $T$ or $P$. 
However, it should be noted that the intensity of the FSDP (or prepeak) is 
a function not only of the number density of locally favored structures, 
$\bar{S}$, but also their spatial arrangement. 

Finally, we provide other evidence to support 
our explanation. It is widely known that the addition of another 
component into a liquid changes its fragility. 
The famous example is a mixture of SiO$_2$ and Na$_2$O. 
Usually, Na$_2$O is considered to act as a network modifier. 
We propose, more specifically, that Na$_2$O is the breaker of locally 
favored structures (probably, 6-member ring structures in the case 
of SiO$_2$). 
%Na$_2$O destabilizes the locally favored structures. 
This concept naturally indicates that the addition of Na$_2$O 
increases the fragility of liquids, and weakens the boson peak 
\cite{Chemarin}, the low-temperature anomaly 
of $C_P$ \cite{White}, and the FSDP, since it induces 
the reduction of the number density of locally favored structures, 
$\bar{S}$. 
The same scenario may also be applied to a mixture of 
B$_2$O$_3$ and Li$_2$O \cite{KojimaS}. 

We also mention a recent simulation study 
by Jund et al. \cite{Jund}, which systematically 
changes the strength of directional interactions 
($\Delta E$ in our terminology) for SiO$_2$-type liquids. 
It demonstrated \cite{Jund} that (i) an increase in the strength 
of directional interactions 
increases the intensity of a boson peak and the FSDP and also the number 
density of ring structures (locally favored structures in our terminology) 
and (ii) a liquid with a more medium-range order 
is stronger, or less fragile. 
These findings are all highly consistent with our physical explanation. 

In summary, we propose a physical origin of the boson peak 
on the basis of the two-order-parameter model of liquid: 
Cooperative vibrational modes localized resonantly 
to locally favored structures and their clusters may be 
the origins of the boson peak. 
This model can explain most of the experimental findings, at 
least on a qualitative level. 
We stress that {\it locally favored 
structures exist not only in a supercooled or glassy state of liquid, 
but also in an equilibrium liquid state if $T_{\rm bp}>T_m$}. 
This is consistent with the experimental finding that the boson peak exists 
even above $T_m$ for some strong liquids, which is difficult 
to explain using previous models.

%\vspace{0.5cm}

\end{multicols}

\begin{thebibliography}{99}
\bibitem{note} Throughout this paper, the boson peak intensity 
means that scaled by $\omega[n(\omega)+1]$, where 
$\omega$ is the angular frequency and $n(\omega) 
=[\exp(\hbar \omega/k_{\rm B}T)-1]^{-1}$ is the temperature bose factor, 
where $\hbar$ is Planck's constant. 
It is proportional to the density of vibrational states, $g(\omega)$.
\end{thebibliography}
\end{document}